\documentclass[prb,showpacs,floatfix]{revtex4-1}
\usepackage{graphicx}
\usepackage{color}
\usepackage{amsmath}
\usepackage{dcolumn}
\usepackage{bm}
\newcommand{\fig}[1]{figure~\ref{#1}}
\newcommand{\Fig}[1]{Figure~\ref{#1}}
\newlength{\figwidth}
\newcommand{\tabl}[1]{table~\ref{#1}}

\newlength{\hw}
\newlength{\vvp}
\newlength{\minusspace}
\newcommand{\msp}{\hspace{\minusspace}}
\newlength{\zerospace}
\newenvironment{tablenotes}{\begin{list}{$^{\alph{enumi}}$}
	{\usecounter{enumi}\setlength{\itemsep}{0ex}
	\setlength{\topsep}{1ex}}}
	{\end{list}}
\newcommand{\sstrut}{\rule[-1ex]{0ex}{3.5ex}}
\newcommand{\refjl}[5]{#1 (#5) {\it#2} {\bf #3} #4.\vspace{-1ex}}
\newcommand{\degree}{$^\circ$}
\newcommand{\mub}{$\mu_{\mathrm B}$}
\newcommand{\sinth}{\ensuremath{\sin\theta/\lambda}}
\newcommand{\inA}{\AA$^{-1}$}

\newcommand{\hfo}{HoFeO$_3$}
\newcommand{\hf}{\ensuremath{\frac12}}
\newcommand{\qr}{\ensuremath{\frac14}}
\newcommand{\tqr}{\ensuremath{\frac34}}
\newcommand{\sho}{\ensuremath{_\textup{\scriptsize{{Ho}}}}}
\newcommand{\sfe}{\ensuremath{_\textup{\scriptsize{{Fe}}}}}
\newcommand{\ff}{\textup{\scriptsize{FeFe}}}
\newcommand{\hh}{\textup{\scriptsize{HoHo}}}
\newcommand{\fh}{\textup{\scriptsize{FeHo}}}
\newcommand{\bntwo}{Pbn$2_1$}
\begin{document}
\title{Temperature evolution of magnetic structure of HoFeO$_3$ by single crystal neutron diffraction}
\author{T. Chatterji$^1$,  M. Meven$^2$, and P.J. Brown$^{1,3}$ }
\address{$^1$Institut Laue-Langevin, 71 Avenue des Martyrs, CS 20156 - 38042 GRENOBLE CEDEX 9,France\\
$^2$ Institute of Crystallography, RWTH Aachen University and J\"ulich Centre for Neutron Science (JCNS) at Heinz Maier-Leibnitz Zentrum (MLZ), Lichtenbergstra§e 1, 85747 Garching, Germany\\
$^3$12 Little St Mary's Lane, Cambridge,CB2 1RR, UK
 }

\date{\today}
\begin{abstract}
We have investigated the temperature evolution of the magnetic structures of
HoFeO$_3$ by single crystal neutron diffraction.  The three different magnetic
structures  found as a function of temperature for \hfo\ are described by the
magnetic groups  Pb$'$n$'2_1$, Pbn$2_1$ and Pbn$'2_1'$ and are stable in the
temperature ranges $\approx$ 600-55~K, 55-37~K and 35$>T>2$~K respectively. In
all three the fundamental coupling between the Fe sub-lattices remains the same
and only their orientation and the degree of canting away from the ideal axial
direction varies. The magnetic polarisation of the Ho sub-lattices in these two
higher temperature regions, in which the major components of the Fe moment lie
along $x$ and $y$, is very small. The canting of the moments from
the axial directions is attributed to the antisymmetric interactions
allowed by the crystal symmetry. They include contributions from single ion anisotropy
as well as the Dzyaloshinski antisymmetric exchange. In the low temperature phase two 
further structural transitions are apparent in which the spontaneous magnetisation changes
sign with respect to the underlying antiferromagnetic configuration. In this temperature
range the antisymmetric exchange energy varies rapidly as the
the Ho sub-lattices begin to order. So long as the ordered Ho moments are small the
antisymmetric exchange is
due only to Fe-Fe interactions, but as the degree of Ho order increases the
Fe-Ho interactions take over whilst at the lowest temperatures, when the Ho moments
approach saturation the Ho-Ho interactions dominate. The reversals of the spontaneous
magnetisation found in this study suggest that in
\hfo\ the sums of the Fe-Fe and Ho-Ho antisymmetric interactions have the same sign as one another, but that
of the Ho-Fe terms is opposite.

.\end{abstract}
\maketitle
  \section{Introduction}
 Rare-earth orthoferrites RFeO$_3$ where R is a rare-earth element constitute an important family of magnetic compounds intensively studied over several decades starting from the early forties of the last century. The presence of two distinct magnetic species and the consequent co-existence of magnetic interactions: Fe-Fe, Fe-R, R-R of decreasing strengths make these compounds liable to undergo several magnetic transitions as a function of temperature. Due to the strong Fe-Fe exchange interaction the Fe subsystem usually orders with a N\'eel temperature T$_N \sim 620-740$ K  as a slightly canted antiferromagnetic structure with weak ferromagnetism. Although the interactions leading to the weak ferromagnetism are small they still play an important role in determining the magnetic properties and spin reorientation transitions at lower temperatures. The R-R interaction is relatively weak which means that the rare-earth sub-lattices only order below $T_{NR} \sim 5-10$ K . Above $T_{NR}$ the R ions are paramagnetic but experience the molecular field of the ordered Fe sub-lattice which magnetize them partially \cite{white69,belov76,belov79,treves65,gyorgy68,sirvardiere69,hornreich73,yamaguchi74}. The interactions within the Fe and R sub-systems are distinct from one another, and change with temperature, field and elastic stress; the competition of these with interactions between the two sub-systems lead to several possibilities for spin reorientation transitions which have been studied in both bulk materials and thin films \cite{white69,belov76,belov79,bornfreund95,bombik00,buchel02,bazaliy03}. Recently interest in the rare-earth orthoferrites has been revived by reports of multiferroic properties (co-existence of magnetic and ferroelectric ordering) in some members of the series \cite{tokunaga08,tokunaga09,lee11,kuo14}. Multiferroicity, by which a magnetic field may influence ferroelectric polarisation and electric fields magnetisation, has been reported, although contested, in the three compounds: DyFeO$_3$, GdFeO$_3$ and SmFeO$_3$ \cite{tokunaga08,tokunaga09,lee11,kuo14}. In some materials ferroelectricity appears only on application of magnetic field. \hfo\ is very similar in both structure and magnetic properties to the aforementioned three compounds, so one may conjecture that it will also show similar multiferroic behaviour. It is for this reason that we have undertaken a more detailed study of the temperature variation of its magnetic structure.

The distorted perovskite crystal structure of  \hfo\ \cite{marez:70} is  described using space group Pbmn, its magnetic structures  were determined in early neutron diffraction measurements on powder
samples \cite{koehl:60,mares:69}. It was found that the Fe sub-lattices order  anti-ferromagnetically
in the G-type configuration \cite{wolla:55} with a N\'eel temperature of 647~K. It
was reported that at room temperature the Fe moments are approximately parallel to [100] and at 43~K  lie in a ($1\overline1 0$) plane. Below
a N\'eel temperature of 6.5~K
 the Ho sub-lattices order  with their moments in the $ab$ plane at 27\degree\ to
y leading to a ferromagnetic component of moment parallel to $x$.

So far all investigation on the temperature dependence of the magnetic structure of HoFeO$_3$ were done on powder samples. In this study we have investigated the temperature evolution of the magnetic structure of HoFeO$_3$ using single crystal samples. In the following sections we describe the results of single crystal neutron diffraction experiments and the several magnetic structures and transitions which they imply. Finally we discuss the evolution of the magnetic structure by evoking the antisymmetric Dzyaloshinski interaction. 

The recent observation of a ferroelectric transition at $\approx$225~K \cite{bulk} means that the true point symmetry of \hfo\ below this transition cannot be higher than mm2, making the likely space group Pbn2$_1$. In the rest of this paper the
space group  Pbn2$_1$ is assumed in all the structure refinements and each magnetic 
structure proposed conforms to one of its magnetic sub-groups.

\section{Experimental}
Neutron diffraction measurements have been made on two small single crystals using the hot neutron diffractometer Heidi of the FRM II reactor in Garching, Germany. The crystals were grown by flux method by S.N. Barilo and D.I. Zhigunov in Minsk, Belarus. This diffractomenter provides a beam of monochromatic neutrons with wavelength in the range 0.55 to 1.2 {\AA} . In the present experiment  a wavelength of 0.7925 {\AA} was used which allowed measurement of the intensities of Bragg reflections with \sinth\ up to 1.25 \AA$^{-1}$. The sample was contained in  a closed cycle cryostat with which the temperature could be controlled in the range 2 - 300 K.

\section{Magnetic symmetry and structure factors for \hfo}
 When described using the space group Pbnm both the Fe and Ho atoms occupy special positions.
 The symmetry of the Fe sites is $\bar 1$ and that of the Ho sites $m$. In space group Pbn2$_1$
the Ho and Fe  atoms each occupy a set of 4-fold general positions. The symmetry operations generating the
4 sites are listed in \tabl{ops} and, assuming the atomic shifts associated with the ferroelectric
transition are small, the relative phases and signs of the contributions which each Fe sub-lattice makes to the structure
factor are those listed in \tabl{phasesFe}.
\begin{table}[h]
\caption{Symmetry operations in space group \bntwo\ and atomic positions in the structure of \hfo.}\vspace{.5ex}
\begin{tabular}{ccccccccccc}
\hline
sub-lattice&\multicolumn{4}{c}{Operator}&\multicolumn{3}{c}{Fe Sites}&\multicolumn{3}{c}{Ho sites$^a$}\\
\hline
1&E&    $x$    &    $y$    &    $z$    &0  &\hf& 0 &$x\sho$      &-$y\sho$     &\qr \sstrut\\
2&b&    $\hf-x$&    $\hf+y$&    $z$    &\hf&0  &0  &\hf-$x\sho$  &\hf-$y\sho$  &\qr \sstrut\\
3&n&    $\hf+x$&    $\hf-y$&    $1/2+z$&\hf&0  &\hf&\hf+$x\sho$  &\hf+$y\sho$  &\tqr\sstrut\\
4&$2_1$&$-x$   &    $-y$   &    $1/2+z$&0  &\hf&\hf&-$x\sho$     &$y\sho$  &\tqr    \sstrut\\[0.5ex]
\hline
\end{tabular}
\begin{tablenotes}
\item $x\sho$=0.98141 and $y\sho$=0.06857
\end{tablenotes}
\label{ops}
\end{table}
\begin{table}[h]
\caption{Relative phases and signs of the contributions of Fe atoms to the structure factors of reflections from \hfo. }
\setlength{\hw}{-1.8ex}
\label{phasesFe}
\begin{tabular}{cllcccccc}
\hline
&&&&&$h+k$&$h+k$&$h+k$&$h+k$\\[\hw]
sub-\\[\hw]
&\multicolumn{3}{c}{Position}&Phase/$\pi$&even&even&odd&odd\\[\hw]
lattice\\[\hw]
&&&&&$l$ even&$l$ odd&$l$ even&$l$ odd\\
\hline
1 &0  &\hf& 0 &0        &+&+&+&+\sstrut\\
2 &\hf&0  &$0$&$h+k$    &+&+&-&-\sstrut\\
3 &\hf &0  &\hf &$h+k+l$&+&-&-&+\sstrut\\
4 &0&\hf&\hf  &$l$      &+&-&+&-\sstrut\\[0.5ex]
\hline
\multicolumn{5}{r}{Reflection Type $^a$}&F&A&C&G\sstrut\\
\hline
\end{tabular}
\begin{tablenotes}
\item The reflection types  have been assigned to match those used by \cite{mares:69}
\end{tablenotes}
\end{table}
The relative phases of contributions from the  Ho sub-lattices
are similar except that, because x\sho\ and y\sho\ are non-zero, cancellation between the positive and
negative contributions is not exact and ordered Ho moments can contribute to all general reflections.
The relative directions of the components of moments of magnetic ions in the 4 sub-lattices  of space group \bntwo\ for the 4 compatible magnetic symmetry groups   are listed in \tabl{signs}.
\begin{table}
\caption{Relative directions of components of moments on magnetic ions on
the different sub-lattices of the \hfo\ structure. }
\vspace{.5ex}
\begin{tabular}{clccccccccccccccccc}
\hline
\multicolumn{2}{c}{sub-lattice}&\multicolumn{3}{c}{ Pbn$2_1\ (\Gamma_1)$}&\quad&\multicolumn{3}{c}{ Pbn$'2_1'\ (\Gamma_2)$}&\quad
&\multicolumn{3}{c}{ Pb$'$n$2_1'\ (\Gamma_3)$}\quad&&\multicolumn{3}{c}{ Pb$'$n$'2_1\ (\Gamma_4)$}\\
\multicolumn{2}{c}{\& Operator}&M$_x$&M$_y$&M$_z$&&M$_x$&M$_y$&M$_z$&&M$_x$&M$_y$&M$_z$&&M$_x$&M$_y$&M$_z$\\
\hline
\ 1 &\  E&   +&+&+&&+&+&+&&+&+&+&&+&+&+\\
\ 2 &\ b$_x$&   +&-&-&&+&-&-&&-&+&+&&-&+&+\\
\ 3 &\ n$_y$&   -&+&-&&+&-&+&&-&+&-&&+&-&+\\
\ 4 &\ $2_{1z}$&-&-&+&&+&+&-&&+&+&-&&-&-&+\\
\hline
\multicolumn{2}{c}{Type}&A&G&C& &F&C&G& &C&F&A& &G&A&F\\
\hline
\end{tabular}
\label{signs}
\end{table}

\begin{figure}[htbp]
   \resizebox{0.6\textwidth}{!}{\includegraphics{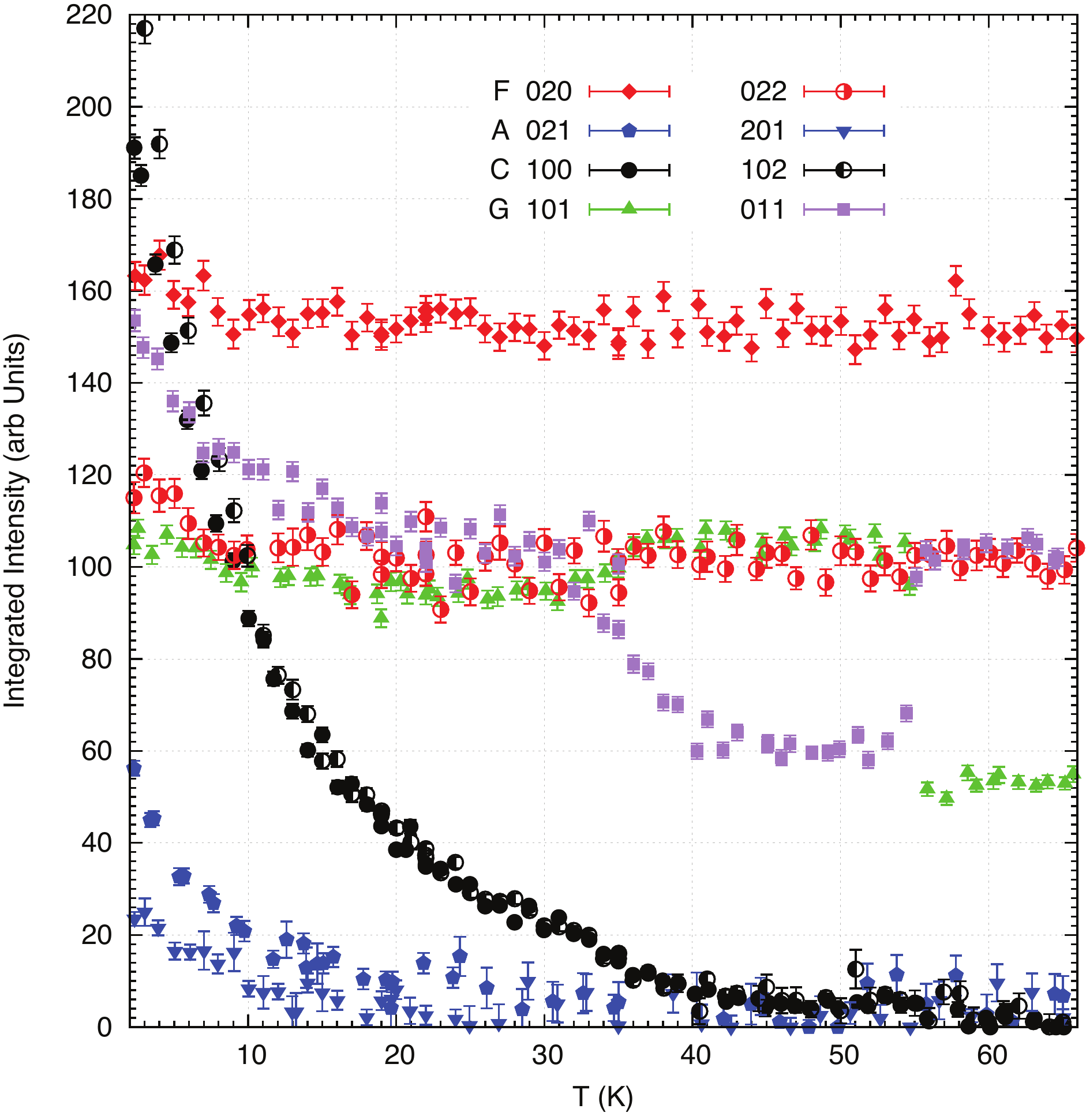}}
   \caption{Temperature dependence of the integrated intensities of selected reflections from an \hfo\ single crystal}
   \label{tdepobs}
\end{figure}
\section{ Temperature dependence of magnetic scattering from \hfo \label{tdep}}
The integrated intensities  of 8 low angle reflections, two from each of the groups F,C,A and G, were
measured from a single crystal of \hfo\ as a function of temperature in the range 2.5 to 65~K
using a neutron wavelength of 0.7925~\AA. The results are presented in \fig{tdepobs}. Above about 40~K significant
 intensity is only observed in the 020 and 022 nuclear reflections (type F) and in 101 and 011 (type G)
 reflections. The latter are systematic absences for nuclear scattering in \bntwo\ and must be due to magnetic scattering from Fe. Inspection of \tabl{signs} shows that G type reflections can occur for
three out of the 4 magnetic groups. For the magnetic group Pbn2$_1$ there must be components of moment parallel
to $y$, for Pbn$'2'_1$ to $z$  and for  Pb$'$n$'2_1$ to $x$.
The observation that 011 is significantly more intense than
101 supports the conclusion \cite{koehl:60} that in this temperature range the Fe moments are nearly parallel to $x$, the magnetic group is therefore  Pb$'$n$'2_1$. At $\approx$ 55~K \fig{tdepobs} there is a sharp transition in which the relative intensities of 011 and 101 change over. Between 55 and 45~K  011 is significantly weaker than 101 although there is no notable change in any of the other reflection intensities.
This behaviour confirms that there is a  reorientation transition at $\approx$ 55~K  in which the magnetic
 space group changes to Pbn2$_1$ and the Fe moments switch to the $y$ direction. Between 40 and 30~K the intensity of the 011 reflection gradually recovers, whilst that of 101 drops very slightly. This suggests
 a further change in magnetic space group  to Pbn$'2'_1$ with the major component of the Fe moments reorienting  parallel
 to $z$. In this temperature range also intensity starts to appear in the 100 and 102 C type reflections
 and the intensity of these reflections increases rapidly at lower temperatures in a way which suggests that it is due to ordering of
 the Ho sub-lattices. In the magnetic group Pbn$'2'_1$ the C type reflections are due to $y$ components
 of moment so it can be concluded that when the Ho sub-lattices order their major moment component is
  in the $y$ direction.
\section{Determination of the  magnetic structure parameters}
\subsection{The 65~K structure: absolute scale and extinction}
\label{Extinc}
With the polar space group \bntwo\  both the Fe and Ho occupy general positions so that, although the magnetic space group  dictates the relative directions
of the components of moment on the different sub-lattices, there are no symmetry constraints on the magnitude and direction of the moments at the sites defined by the identity operator E.
To determine these parameters the integrated intensities of a large number of reflections (911)
with $\sinth<1.25$~\inA\ have been  measured at 65~K, 35~K and 2.5~K. The 65~K data were
used to confirm the positional parameters, obtain a model
 for extinction in the crystal and determine the absolute scale of the measured intensities.
Previous analysis of more limited data measured at a longer wavelength (2.36 \AA) had shown that extinction was severe and that the the extinction and scale
parameters were highly correlated.
The early work on \hfo\ \cite{koehl:60} and our measurements at 2.36 \AA\
 at temperatures up to 300~K, show that at 65 K the Fe sub-lattices  are almost fully ordered and that the magnetic scattering is confined to the type G reflections (those with $h+k$ and $l$ both odd). Based
 on this and on the conclusions of section \ref{tdep} it was assumed that the magnetic scattering at 65~K could be modelled using the magnetic space group Pb$'$n$'2_1$ with Fe moments of 4.6 \mub\ oriented parallel to $x$.
 Least squares refinements of the nuclear positional
 parameters, the scale factor and a single extinction parameter were carried out using this magnetic model.
 For refinement of the positional and temperature parameters the centrosymmetric space group Pbnm was used,
 the refined parameters are listed in \tabl{refpars}.  Again there was a high (90\%) correlation between the scale factor and the extinction parameter, but reasonable agreement between observed and calculated structure factors  was obtained with an R factor of 8.3 on structure factors and a goodness of fit $\chi^2=3.8$. The relatively high value of the residual R factor is probably due to imperfect modelling of the extinction. The model predicted that extinction reduces the intensities of the strongest (nuclear) reflections by a factor of nearly 3 and the strongest magnetic ones by factors of up to 2. It may be noted that the only really
 significant anisotropic temperature coefficients in \tabl{refpars} are the B(11) of iron and both oxygens which suggests that the principal
 atomic shifts in the ferroelectric transition are in the $x$ direction.

\begin{table}[htp]
\caption{Crystal structure parameters for \hfo\ obtained from least squares refinement of integrated intensity
measurements at 65~K}
\begin{center}
\setlength{\hw}{-1.5ex}
\setlength{\vvp}{0.8ex}
\begin{tabular}{cllllll}
\hline
& & & &  \multicolumn{1}{c}{B(11)$^a$} &\multicolumn{1}{c}{B(22)} &\multicolumn{1}{c}{B(33)}\\[\hw]
Atom    &\multicolumn{1}{c}{$x$} &\multicolumn{1}{c}{$y$} &\multicolumn{1}{c}{$z$}\\[\hw]
& & & &  \multicolumn{1}{c}{B(23)} &\multicolumn{1}{c}{B(13)} &\multicolumn{1}{c}{B(12)}\\[\vvp]
\hline & & & &                                 \msp0.03(3) & -0.03(3)      &  -0.06(3)\\[\hw]
Ho &   0.9813(2)&  0.0686(2) &  0.2500 \\[\hw]
 & & & &                                             &               &  -0.00(3) \\[\vvp]
 & & & &                                 \msp0.11(2) & -0.02(3)      &  -0.04(2)\\[\hw]
  Fe &     0      &  0.5000     &0\\[\hw]
 & & & &                                 \msp0.04(2) & -0.00(2)      &  -0.02(2)\\[\vvp]
 & & & &                                 \msp0.13(5) & \msp0.06(4)   &  -0.10(4)\\[\hw]
  O1  &  0.1098(3)&  0.4601(3) &0.2500 \\[\hw]
 & & & &                                             &               & -0.02(4) \\[\vvp]
 & & & &                                \msp0.12(3)  &  -0.04(3) & 0.02(3)\\[\hw]
  O2  &  0.6920(2)  &0.3048(2)    &0.0564(2)\\[\hw]
 & & & &                                -0.04(3)     & \msp0.01(3)   & -0.04(3)\\[\vvp]
 \hline
\end{tabular}
\begin{tablenotes}
\item The anisotropic temperature parameters B(ij) are given in units of \AA$^{-2}$.
\end{tablenotes}
\end{center}
\label{refpars}
\end{table}%
\subsection{Magnetic structure at 35~K and 2.5 ~K}
The arguments presented in section~\ref{tdep} make it likely that at 35~K the magnetic structure is
transforming between configurations with different magnetic space groups:   Pbn$2_1$ and Pbn$'2'_1$.
Least squares refinements of the magnetic parameters were carried out for both groups using the crystal structure
parameters, extinction  and scale determined from the 65~K data and assuming the same saturated Fe moment
of 4.6~\mub. The results are presented in \tabl{compt}. A marginally better agreement was obtained with
Pbn$'2'_1$ although it is probable that at this temperature some intermediate or mixed phase state is present.

\begin{table}[htp] {
\footnotesize\setlength{\hw}{-1.2ex}

\caption{Components of the magnetic moments of Fe and Ho \hfo\ obtained from least squares refinements
using data collected at 35~K and 2.5~K}
\vspace{0.5ex}
\begin{tabular}{clllllllcc}
\hline
&\multicolumn{1}{c}{Magnetic}&\multicolumn{3}{c}{Fe (\mub)}&\multicolumn{3}{c}{Ho (\mub)}\\[\hw]
\multicolumn{1}{c}{T (K)}&&&&&&&&$^aR_{\textup{\scriptsize{cryst}}}$&$\chi^2$\\[\hw]
&\multicolumn{1}{c}{Group}&\multicolumn{1}{c}{$M_x$}&\multicolumn{1}{c}{$M_y$}&\multicolumn{1}{c}{$M_z$}
&\multicolumn{1}{c}{$M_x$}&\multicolumn{1}{c}{$M_y$}&\multicolumn{1}{c}{$M_z$}\\[\vvp]
\hline
&{Pbn$2_1$}&      \msp1.0(2) & -4.45(5) & -0.54(6) &  -1.1(2) &      \msp0.7(2) &      \msp1.64(15)&
9.47 &4.5 \\[\hw]
35\\[\hw]
&Pbn$'2'$&      \msp0.6(3)  & -0.08(6)  &    -4.55(4)&\msp1.1(2)  &       \msp1.2(2) &   -0.9(2)&
 9.37 &3.8\\
 \hline
2.5& Pbn$'2'$&  \msp0.47(14)& -0.24(12) & -4.57(2)&  \msp3.53(7) &   \msp7.90(6)   &  \msp0.1(7)& 7.4 &4.2 \sstrut\\
\hline
\end{tabular}
\begin{tablenotes}
\item $R_{\textup{\scriptsize{cryst}}}=\sum_i|F_{\textup{\scriptsize{obs}}}-F_{\textup{\scriptsize{calc}}}|/
\sum_i|F_{\textup{\scriptsize{obs}}}|$
\end{tablenotes}
\label{compt}
}\end{table}%
\subsection{Temperature variation of the magnetic parameters}
\begin{figure}[htbp]
\resizebox{\textwidth}{!}{\includegraphics{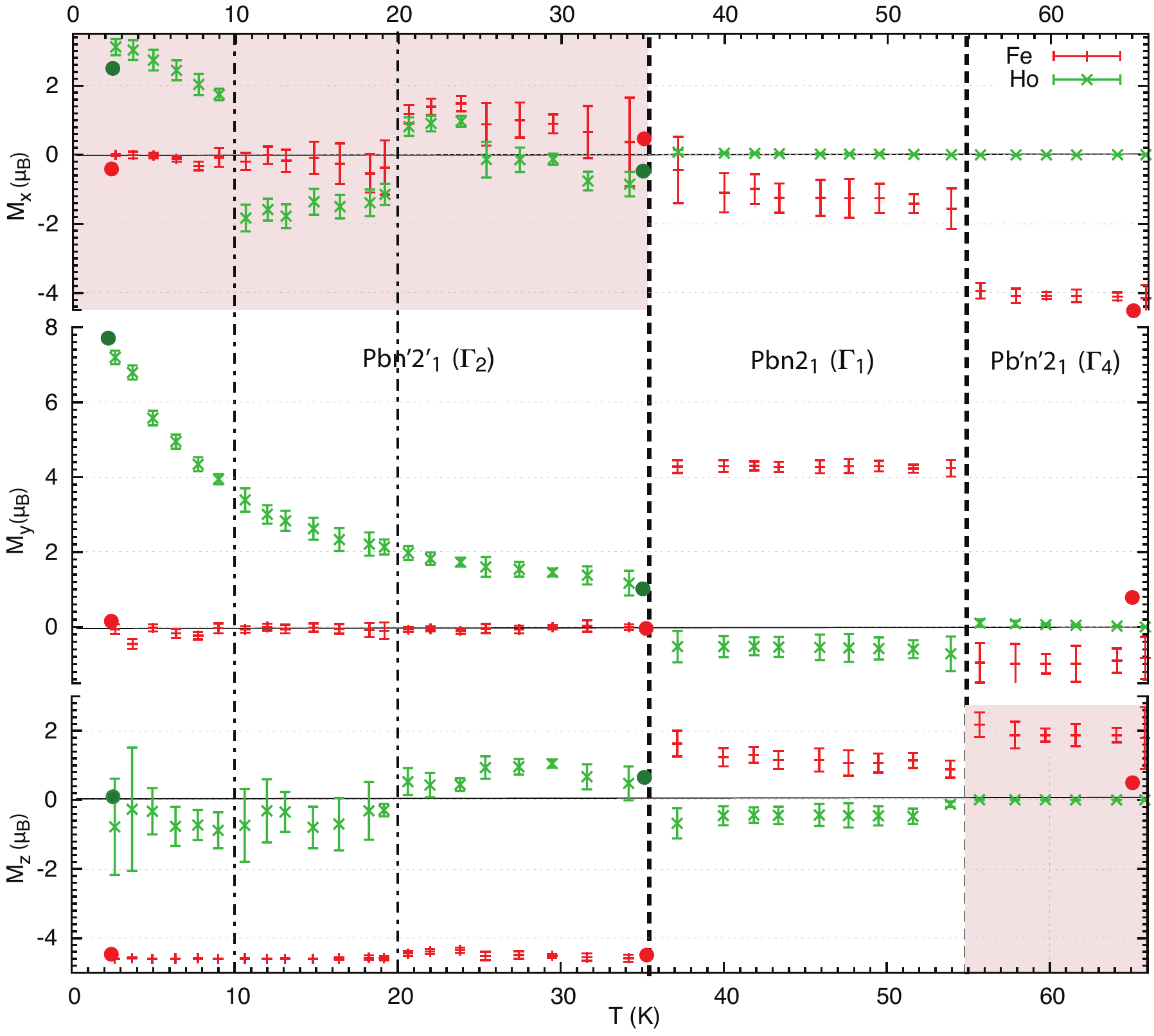}}
\caption{Temperature variation of the x,y, and z components of moment on Fe and Ho ions in \hfo\ in the temperature
range 2.5 to 66~K determined from 8 low angle reflections. Values obtained from the three large data sets are shown as filled circles. The shaded areas mark the regions where the components
on the 4 sub-lattices are ferromagnetically aligned}
\label{comps}
\end{figure}

In order to trace how the details of the magnetic structure change with temperature the measurements of
the 8 reflections followed over the whole temperature range 2 - 65~K were analysed further. The data were separated
into groups for which measurements of all 8 reflections were present at temperatures within a temperature range of 2.5 ~K.
For each group the values of 5 parameters were refined, these were the magnitude and orientation of the Ho moment and just the orientation of the Fe moment its magnitude being fixed at 4.6\mub. The magnetic space groups used in different temperature ranges were:
66 - 55K  Pb$'$n$2_1'$, 55 - 37~K Pbn$2_1$  and 34 - 2.5~K Pbn$'2_1'$. The temperature dependence of the x,y,and z components of moment on the representative (E) sub-lattice for Fe and Ho ions obtained is shown in \fig{comps}.
Besides the anomalies round 55~K and 35~K associated with the reorientation transitions there appears to be a significant
break in the variation of the x components at 20~K and again at 10~K. In this temperature range the x-components of
moments are aligned ferromagnetically and the breaks are associated with abrupt changes in the direction in which the Ho moment is canted away from the y-axis.

\section{Discussion}
\begin{figure}[htbp]
\resizebox{0.5\textwidth}{!}{\includegraphics{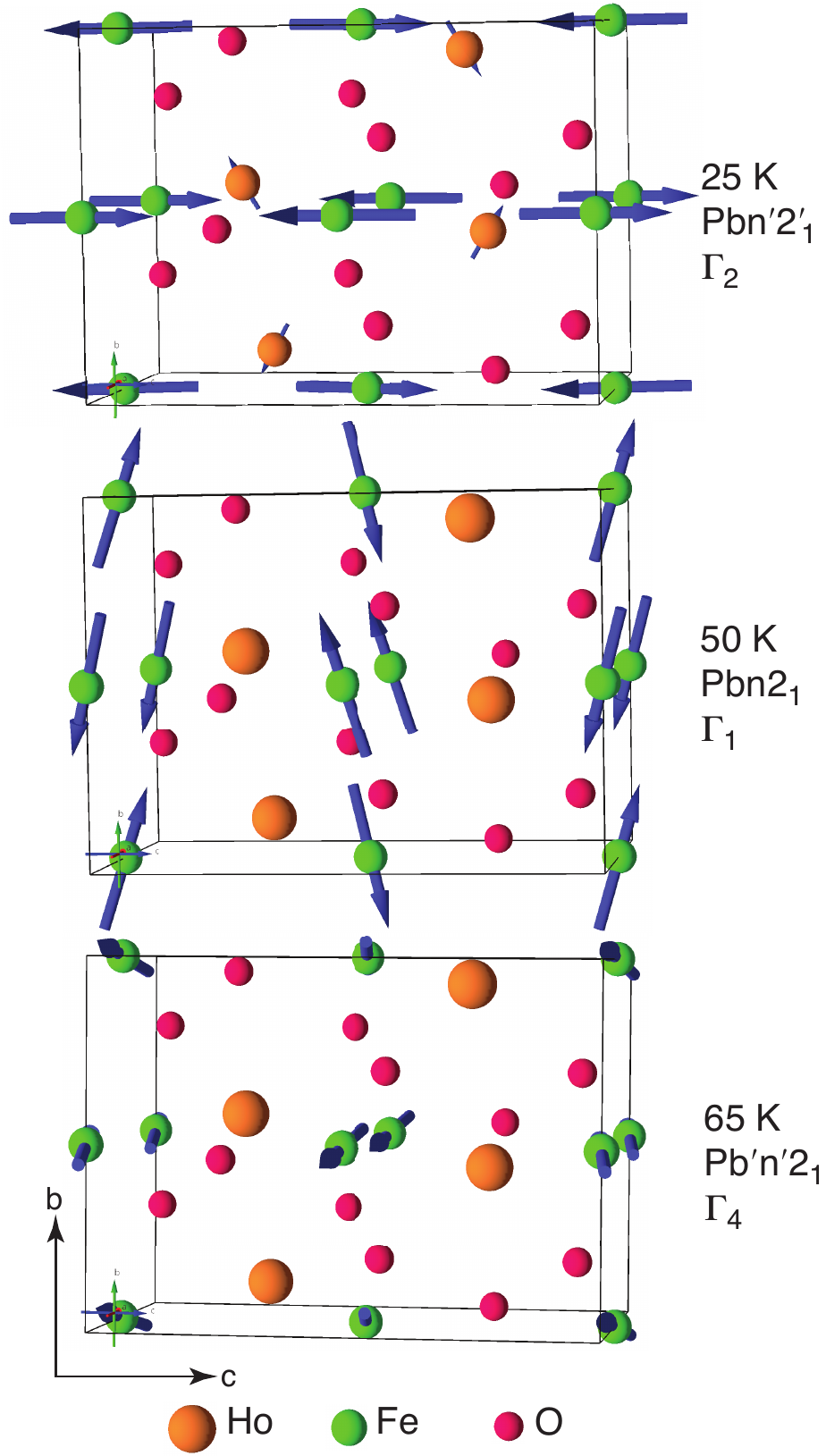}}
\caption{Schematic representation of the three modulation modes found for the ordered magnetic structures of \hfo.}
\label {modelc}
\end{figure}
	The three different magnetic structures  found as a function of temperature for \hfo\ are illustrated schematically in
\fig{modelc}.
They are described by the magnetic groups  Pb$'$n$'2_1$, Pbn$2_1$ and Pbn$'2_1'$ and are stable
in the temperature ranges $\approx$ 600-55~K, 55-37~K and 35$>T>2$~K respectively. In all three the fundamental coupling between the Fe moments remains the same and only their orientation and the degree of canting away from
the ideal axial direction varies. The magnetic polarisation of the Ho
sub-lattices in these two higher temperature regions, in which the major components of the Fe moment lie along $x$
and $y$, is very small.
The origins and type of the reorientation transitions which can occur in the rare earth orthoferrites have been discussed by Yamaguchi \cite{yamaguchi74}. He shows that an abrupt transition from $\Gamma_4$ (Pb'n$'2_1$) to $\Gamma_1$  (Pbn$2_1$), such as that at 55~K  in \hfo, can occur when, with falling temperature,  the energy due to anisotropic exchange between Ho and Fe ions outweighs that due to Fe anisotropy  alone which favours the $\Gamma_4$ configuration. He does not include criteria for the non-abrupt $\Gamma_2$ to $\Gamma_1$ transition  which
takes place in \hfo\ between 37 and 30~K.

\begin{figure}[htbp].
\resizebox{\textwidth}{!}{\includegraphics{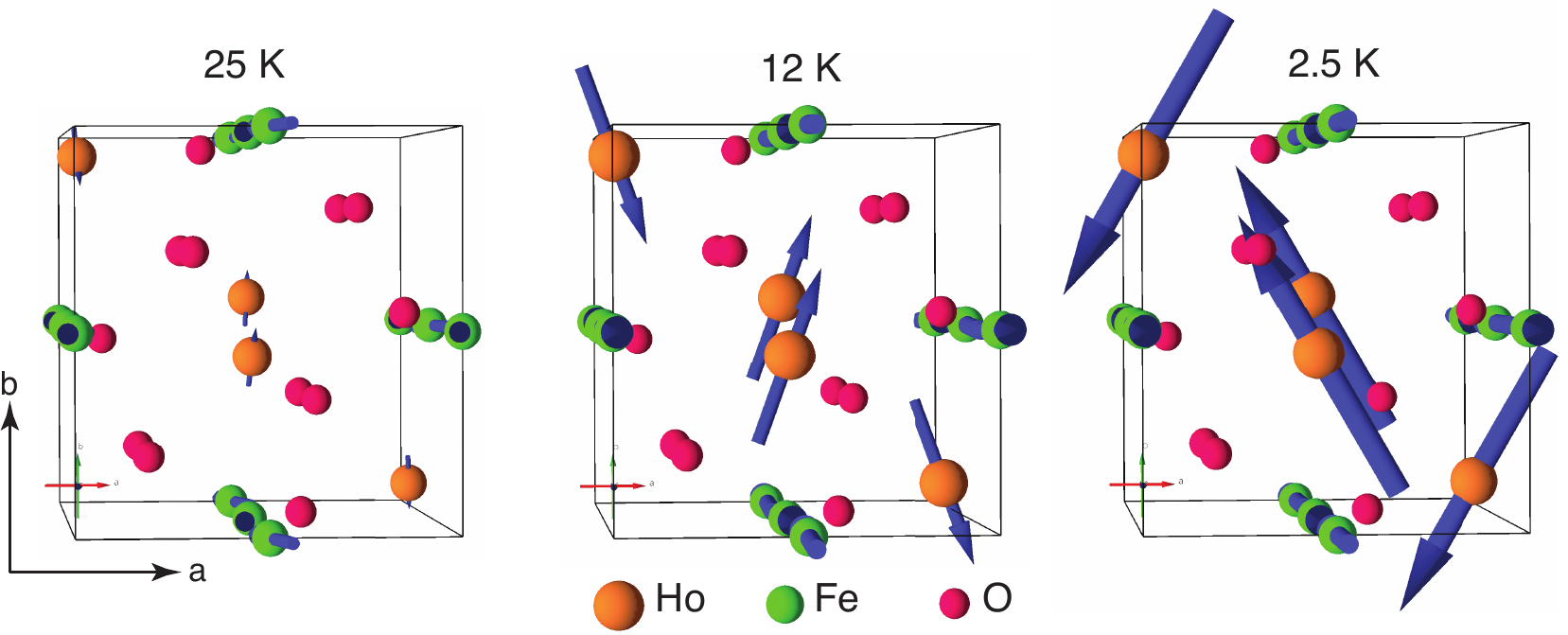}}
\caption{Variation of the relative orientations of the Fe and Ho moments in \hfo with increasing order in the Ho sub-lattices.}
\label {model_cold}
\end{figure}

\Fig{model_cold} illustrates the evolution of the structure as the Ho sub-lattices order.
The major component of the Ho moment is always parallel to $y$ and the anisotropic exchange interaction forces the Fe moments to to reorient along $z$. In Pbn$'2_1'$ components
of moment in the $x$ direction are ferromagnetically aligned and it is the direction of this ferromagnetic
moment relative to  ${\bf M}\sho \times {\bf M}\sfe$   which  changes sign twice on cooling from 30 - 2.5~K. 
A similar reversal of the spontaneous ferromagnetic moment with respect to the underlying antiferromagnetic structure
has been observed in SmFeO$_3$ \cite{lee11} although in that case only a single reversal at 5~K, rather than the 
double reversal indicated here, was found.

Weak ferromagnetism in the orthoferrites has been attributed to to the antisymmetric exchange interaction between magnetic ions \cite{gorod:64}. In \hfo\ below 30~K it
is the sum of contributions from both Fe and Ho and will have the form
\begin{align}
 E_\textup{\scriptsize{{antisymmetric}}} = &\sum_{i=1,4}^{j=i+1,4} {\bf D}_{\ff,i,j}\cdot({\bf M}\sfe(i)\times{\bf M}\sfe(j))
  + {\bf D}_{\hh,i,j}\cdot({\bf M}\sho(i)\times{\bf M}\sho(j))\nonumber\\
 &+\sum_{i=1,4}^{j=1,4}({\bf D}_{\fh,i,j}\cdot{\bf M}(\sfe(i)\times{\bf M}\sho(j))
\label{antisym}
\end{align}
The ${\bf D}_{n,i,j}$ are Dzyaloshinski vectors \cite{dzyal:58} which may include contributions from single ion anisotropy as well as
antisymmetric exchange. This antisymmetric exchange energy varies rapidly in the temperature range in which the Ho sub-lattices
are ordering. Whilst the ordered Ho moments are small it is dominated by the terms containing ${\bf D}_{\ff}$ and the sense
of the ferromagnetic moment is determined by the first summation in eqn.\ref{antisym}. As the degree of Ho order increases
the terms in the third summation, that containing products of the Fe and Ho moments, begin to exceed those dependent on Fe moments
only and the sense of the ferromagnetic moment may change. Finally at the lowest temperatures when the Ho moments approach saturation
the Ho-Ho interactions dominate. The behaviour observed in this experiment shows that for \hfo\ the sums of the Fe-Fe and Ho-Ho
interactions have the same sign as one another, but that of the Ho-Fe terms is opposite.

\section{Possible ferroelectricity in HoFeO$_3$}
We mentioned already that there is an unpublished report of ferroelectricity \cite{bulk} on HoFeO$_3$ that motivated us to reinvestigate the temperature evolution of magnetic structure of HoFeO$_3$ carefully by single crystal neutron diffraction. The report of ferroelectricity is based on pyrocurrent measurements on a single crystal sample and is also supported by synchrotron single-crystal X-ray diffraction experiments. The ferroelectric transition temperature was determined to be around $T_C \approx 225$ K. But recently some doubt has arisen also as to the reproducibility of the ferro-electric transition reported to occur \cite{lee11} at 
the N\'eel temperature in SmFeO$_3$ and it is argued\cite{kuo14} that the inverse Dzyaloshinski interaction cannot drive
a ferroelectric transition in a $k = 0$ antiferromagnet. 
The ferroelectric transition observed in HoFeO$_3$
is reported\cite{bulk} to occur at 225 K well below the N\'eel temperature for Fe order and in a $ k = 0$ structure so
the ferroelectric transition must be driven by some mechanism other than the inverse DM effect. 
However the loss of centrosymmetry associated with such a transition removes the symmetry 
constraint which compels the Ho moment
to be either parallel or perpendicular to the a-b plane, and allows Ho to participate in 
weak z-axis ferromagnetism at low temperature. This ferromagnetism would be expected to couple strongly
to the ferro-electric polarity. 

\section{Acknowledgement}
This work is based upon experiments performed at the Heidi instrument operated by JCNS at the Heinz Maier-Leibnitz Zentrum (MLZ), Garching, Germany. The author TC gratefully acknowledges the financial support provided by JCNS to perform the neutron scattering measurements at the Heinz Maier-Leibnitz Zentrum (MLZ), Garching, Germany. TC wishes to thank S.N. Barilo for suppling the HoFeO$_3$ single crystals used in the present study.


\begin{thebibliography}{99}
\bibitem{white69}R. White, J. Appl. Phys. {\bf 40}, 1061 (1969).
\bibitem{belov76}K.P. Belov, Sov. Phys. Usp {\bf 19}, 574 (1976).
\bibitem{belov79} K.P. Belov, A.K. Zvezdin, A.M. Katomtseva, and R.Z. Levitin, \emph{Orientation Phase Transitions in Rare Earth Magnetic Materials} (Nauka, Moscow, 1079).
\bibitem{treves65} D. Treves, J. Appl. Phys. {\bf 36}, 1033 (1965).
\bibitem{gyorgy68} E.M. Gyorgy, J.P. Remeika, And F.B. Hagedorn, J. Appl. Phys. {\bf 39}, 1369 (1968).
\bibitem{sirvardiere69} J. Sirvardiere, J. Solid State Comm. {\bf 7}, 1555 (1969).
\bibitem{hornreich73}R.M. Hornreichand I. Yaeger, I. J. Magn. {\bf 4}, 71 (1973).
\bibitem{yamaguchi74} T. Yamaguchi, J. Phys. Chem. Solids {\bf 35}, 479 (1974).
\bibitem{bornfreund95}R.E. Bornfreund et al., J. Magn. Magn. Mater. {\bf 151}, 181 (1995).
\bibitem{bombik00}A. Bombik and A.W. Pacyna, J. Magn. Magn. Mater. {\bf 220}, 18 (2000).
\bibitem{buchel02}V.D. Buchel\,nikov et al., JETP {\bf 122}, 122 (2002).
\bibitem{bazaliy03}Ya. B. Bazaliy et al., Physica B {\bf 329-333}, 1257 (2003).
\bibitem{tokunaga08}Y. Tokunaga, S. Iguchi, T. Arima and Y. Tokura, Phys. Rev. Lett. {\bf 101}, 097205 (2008).
\bibitem{tokunaga09}Y. Tokunaga, N. Furukawa, H. Sakai, Y. Taguchi, T. Arima and Y. Tokura, Nat. Mat.  {\bf 8}, 558 (2009).
\bibitem{lee11}J.-H. Lee, Y.K. Jeong, J.H. Park, M. Oak, H. M. Jang, J.Y. Son and J.F.  Scott, Phys. Rev. Lett {\bf 107}, 117201 (2011).
\bibitem{kuo14}C. -Y. Kuo et al., Phys. Rev. Lett. {\bf 113}, 217203 (2014).
\bibitem{marez:70}M. Marezio, J.P. Remeika and P.D. Dernier, Acta Cryst.B {\bf 26},2008, (1970)
\bibitem{koehl:60}W.C. Koehler,  E.O. Wollan and M.K. Wilkinson, Phys. Rev.{\bf 118}, 58 (1960).
\bibitem{mares:69} J. Mareschal and J. Sivardi\`ere J, J.de Physique {\bf 30}, 967 (1969).
\bibitem{wolla:55} E.O. Wollan and W.C. Koehler, Phys. Rev {\bf 100}, 545 (1955).
\bibitem{bulk} S. Giri et al. (to be published)
\bibitem{gorod:64}G. Gorodetsky G and D. Treves , Phys. Rev {|bf 135}, A97 (1964).
\bibitem{dzyal:58} I.E. Dzyaloshinsky, J. Phys. Chem. Solids{\bf 4}, 241 (1958).
\end{thebibliography}
\end{document}